# SONIFICATION AESTHETICS AND LISTENING FOR NETWORK SITUATIONAL AWARENESS

*Paul Vickers, Christopher Laing, Mohamed Debashi*

Northumbria University, Dept. of Computer Science & Digital Technologies, Pandon Building Camden St, Newcastle upon Tyne, NE2 1XE, UK
{paul.vickers, christopher.laing, mohamed.debashi}@northumbria.ac.uk

*Tom Fairfax*

SRM Solutions, The Grainger Suite, Dobson House, Regent Centre, Gosforth, Newcastle upon Tyne, NE3 3PF, UK
tom.fairfax@srm-solutions.com

## ABSTRACT

This paper looks at the problem of using sonification to enable network administrators to maintaining situational awareness about their network environment. Network environments generate a lot of data and the need for continuous monitoring means that sonification systems must be designed in such a way as to maximise acceptance while minimising annoyance and listener fatigue. It will be argued that solutions based on the concept of the soundscape offer an ecological advantage over other sonification designs.

## 1. INTRODUCTION

In military circles there is debate about whether cyberspace has become the fifth warfighting domain (the others being sea, land, air, and space) [1]. Computer networks are increasingly coming under strain both from adversarial attacks (warfighting in military parlance) and from load and traffic pressures (e.g., increased demand on web services). Another term that has made its way from the military lexicon into the wider world of network administration is *situational awareness*. Endsley [2, p. 36] defined situational awareness (SA) as the

> …perception of elements in the environment within a volume of time and space, the comprehension of their meaning, and the projection of their status in the near future.

So, SA facilitates an administrator in becoming aware of a network's current state. The perception phase of SA comprises the recognition of situational events and their subsequent identification. Sonification is a process of computational perceptualisation which Vickers [3] suggested is well suited to the monitoring of time-dependent processes and phenomena such as computer networks. Vickers, Laing, and Fairfax [4] proposed a soundscape-based method for sonifying computationally-derived properties of network traffic in order to monitor in real time the stresses being experienced by a network. Initial results were promising but an important design challenge involving design aesthetics remains to be addressed.

## 2. THE CYBER ENVIRONMENT

It has been noted that there is debate in military circles about whether cyberspace has become the fifth warfighting domain (following land, sea, air, and space). The central focus of debate is the cyber environment (sometimes known as cyberspace) is a discrete area of operations or whether it is a more pervasive concept that runs through all of the other domains. While land, sea, air, and space are physically distinct and are defined by similar criteria, cyberspace is defined in a different way, existing on an electronic plane rather than a physical and chemical one. Some argue that cyberspace is merely a common component of the four other domains rather than a discrete domain of its own. Indeed, it is easy to see how cyber operations can play a significant role in land, sea, air or space warfare, due to the technology employed in each of these domains [1].

This distinction depends on the way that the various domains are defined. If our definitions are underpinned by a purely physical paradigm, then it is arguable that cyberspace is a very different type of context to the traditional warfighting domains. If, however, our definitions are based on an operational paradigm, then the distinction is less clear. It is possible to conduct entire operations in the cyber environment, made possible by the interconnected nature of the Internet and associated infrastructures. In the same way, it is common to have joint operations operating across multiple domains, including the cyber environment, and the cyber environment isn't restricted to military warfighting scenarios.

Though operations in cyberspace are complex, they can be simplified, to some extent, by the cyber operations spectrum. This divides cyber operations into 3 areas [1]:

- **Defence**: Defensive operations take up approximately 80% of cyber activity. This constitutes the work that is (or should be) undertaken by all individuals or organisations. It ranges from simple protection of individual personal equipment to complex security management architectures.
- **Exploitation**: Exploitation is covert activity conducted within an adversaries area of operations. This is generally invisible to the defender (unless compromised by the defender). Exploitation operations range from preparatory activity conducted to enable future activity to protracted information farming operations which are designed to generate intelligence over a protracted period of time.
- **Attack**: The overt phase when effect is brought to bear on a target. There are a wide range of exploits and strategies associated with this phase. It should be noted that a visible attack may well have been preceded by invisible exploitation operations.

A knowledge of where current operations lie within the cyber spectrum is critical to a clear understanding of the cyber environment. It is also helpful to view the actions of adversaries in this context in order to try to understand the adversarial plan and predict their likely future actions.

Traditional protective strategies were often based on the defence of boundaries and perimeters. Whether defended by technology or, in some cases, complete air gaps, boundary based defence was initially effective until attackers found ways to achieve a breach, whether by compromising vulnerable technology or bridging air gaps, as could be seen, for example, in the Stuxnet attack on the Iranian nuclear processing facility. This boundary-based model is increasingly seen as flawed due to the enormous complexity and granularity of the cyber environment. Increasingly, defensive architectures are seen to be resilient matrices of multiple defensive components. It is no longer credible for organisations to assume that they are completely safe. The sensible security strategy now focuses on raising the bar to reduce the likelihood of a successful attack, but to assume that a proportion of attacks will be successful, but to have the mechanisms in place to identify and manage these events when they occur. Organisations must also ensure that operational architectures are sufficiently resilient to enable them to continue to operate whilst 'under fire' and to be able to accept known levels of attrition. This has resulted in a subtle but tangible shift from purely protective postures to proactive intelligence management within organisations.

In many cases, the compromise of technology is achieved indirectly. This often involves the compromise of people. A wide range of social engineering attacks are employed in order to compromise technology using traditional human weaknesses, including greed, curiosity, insecurity and ignorance. The dependence of cyberspace on people also extends the scope of compromise from direct attacks on target systems, to indirect targeting of social, economic, commercial and financial architectures. The traditional 'high threat club' (those organisations who are known to represent high value targets to attackers) are no longer the only organisations with a requirement for active and dynamic information security infrastructures. Information security is now a critical aspect of corporate governance across the organisational spectrum.

An important driver for the cyber environment is that it effectively becomes an asymmetric enabler. Cyber operations provide a viable attack vector for small nations or influence groups that enables them to directly engage even the largest power bases (military or otherwise) worldwide. One of the effects of the advent of the cyber environment has been to remove much of what von Clausewitz (1873) termed the friction of war. This is exacerbated by the fact that tempo changes are possible, where operations can move rapidly from slow, covert activity to high intensity attack activity with little physical impact.

History has shown that an ability to switch tempo in battle has enormous value in its ability to unhinge adversaries and to compromise their will and ability to fight. This is one of the characteristics that lies at the heart of the 'manoeuverist' doctrine that underpins much of the 20th century warfighting doctrine. Manoeuver warfare is a potentially complex doctrine which is built on simple principles which shape the chosen battlefield through knowledge, understanding and agility. The British Army describes the manoeuverist approach as follows [5]:

> This is an indirect approach which emphasises understanding and targeting the conceptual and moral components of an adversary's fighting power as well as attacking the physical component. Influencing perceptions and breaking or protecting cohesion and will are essential. The approach involves using and threatening to use force in combinations of violent and non-violent means. It concentrates on seizing the initiative and applying strength against weakness and vulnerability, while protecting the same on our own side. The contemporary Manoeuvrist Approach requires a certain attitude of mind, practical knowledge and a philosophy of command that promotes initiative. (Chapter 5)

The cyber environment provides an additional dimension within which agility can be achieved, and initiative seized. It is, perhaps, instructive that the practical application of the manoeuverist approach is broken down into the following components:

- **Understanding the situation**: using information, intelligence and intuition coupled with a sound understanding of objectives and desired outcomes.
- **Influencing perceptions**: planning, gaining and maintaining influence, and the management of key stakeholders.
- **Seizing and holding the initiative**: Ensuring that we hold the ability to dictate the course of events, through competitive advantage, awareness and anticipation.
- **Breaking cohesion and will in our adversaries**: Preventing our adversaries from being able to co-ordinate actions effectively, and compromise their determination to persist.
- **Protecting cohesion and will in ourselves and our allies**: Enabling our own freedom of action and ability to co-ordinate our resources, ensuring that we retain the will and coherence to operate.
- **Enhancing and evolving the approach through innovation**: The approach is enhanced through simplicity, flexibility, tempo, momentum and simultaneity.

All of these components are areas where cyber operations can play a significant part both for the attacker and the defender. In military terms, cyber may be seen as a force multiplier, increasing the effect of existing operational capability. There is, however, another side, in that these principles and components can be applied to operations in the cyber environment and, if applied with flexibility, can provide structure to planning.

Cyberspace is characterised, amongst many things, by a lack of natural visibility and tangibility. Humans have sense-based defensive postures. Sight, smell, feel and sound underpin our innate defensive posture. The challenge of cyberspace is that none of these senses, the core of our sensory toolkits, are effective in the cyber environment without technology and tools. It could be said that we have created an operating environment for which we do not yet have effective sensory perception. We therefore become dependent on these tools, and the way in which they have been developed and configured. This inability to engage our senses in a native manner represents an opportunity for attackers and defenders. In this environment, clear understanding of the current state of the battlespace; situational awareness, becomes a battle winning factor.

To return to the question — has cyber become the new battle space? — whilst the role of the cyber environment as a fully-fledged warfighting domain is open to sustained debate, it is very clear that the cyber environment is one in which it is possible to conduct a range of targeted operations. It is also clear that these operations may be conducted in isolation, or in conjunction with operations in the kinetic sphere (in any of the four principal warfighting domains.)

However we eventually decide to classify this area, we must ensure that we are able to operate within it, at least as effectively as our adversaries are able to. As such, it would be prudent to consider it to be a battlespace, and a high tempo battlespace in which our native situational awareness is limited. It is also a battlespace in which our ability to maintain an agile, proactive posture is critical to our ability to gain and maintain the initiative.

## 3. SITUATIONAL AWARENESS

As outlined above, terms such as 'battlespace' and 'attack' have become common parlance when discussing the protection of information infrastructures from a wide range of cyber-based information operations, as has another term, 'situational awareness'. The study of situational awareness has its roots in military theory [1]. Situational awareness has the goal of understanding the state of a particular scope and using that understanding to make decisions about how to proceed and respond to events. There are different models and frameworks for situational awareness in the computer networks field, but there is general agreement that at its core lie three levels of awareness (see Endsley [2]):

1. **Perception**: becoming aware of situational events;
2. **Comprehension**: interpreting what is happening to form a situational understanding of the events;
3. **Projection** (i.e., prediction): using the understanding to inform what actions (if any) should be taken to control the network.

When discussing the manoeuverist approach above, we noted that in order to gain and maintain the initiative in a particular area of operations, the first step or component was to achieve an understanding of the area and activity within it. This clearly echoes Endsley's model noting a perception and comprehension of information in order to enable projection; actions to seize the initiative in a particular situation [1].

Noting that the manoeuverist perspective on situational awareness developed within a kinetic warfighting context [5], it looks in even more detail at information operations, intelligence collection and collation as part of the process to convert perception to comprehension and projection. This is directly relevant to the information space and implies a degree of planning and direction through the acquisition, analysis and dissemination of intelligence. In many contexts, analysis is intuitive and organic, especially in the high tempo information space, however, we must acknowledge its role as an active part of the practical process. It is this transition from information to intelligence which takes us from Endsley's Understanding Phase to the Projection Phase.

Another practical perspective comes from John Boyd. Whilst Endsley's model is useful for understanding the levels of situational awareness, an example from the kinetic sphere readily illustrates how it adds value in a practical context. If we take a brief step into kinetic military doctrine, and view the computer incident response process in the context of Boyd's OODA loop theory (see Angerman [6]), we find a useful model to review the practical relevance of situational awareness in a combat situation [1].

John Boyd was commissioned by the US Department of Defense in 1976 to analyse why US pilots in Korea were so successful despite the fact that the opposing Chinese MiG–15 aircraft were technically superior in many respects. His simple theory, which postulated that certain aspects of the US aircraft design enabled the pilots to react more quickly to the changing battle, has gained much traction since [1].

Boyd theorised that combat pilots made decisions using a cycle comprising four steps: observe, orient, decide, act (OODA). In a contest between two opposing pilots the individual who could complete this cycle the quickest would have the advantage. Boyd suggested that the increased speed at which the US pilots could react and reorient themselves outweighed the technical superiority of the MiG–15 [1].

Refinements have since been made to Boyd's OODA model and it is particularly pertinent in the context of cyber security and the defence of information networks. The information network environment is characterised by high tempo and granularity, coupled with low visibilty and tangibility. Administrators are therefore dependent on complex and granular data feeds for data about what is happening, and must often further translate this view into language that can be understood by decision makers. The use of tools can simplify this complex data picture, but each analysis layer introduces margin for error and adds Clausewitzian friction. Added to this are the practical limitations of our physical and intellectual physiology; it is practically impossible for most people to sit watching complex visual data feeds concurrently with other activity without quickly losing effectiveness [1].

Network administrators require a real-time monitoring tool to facilitate the acquisition and maintenance of situational awareness. Such a tool would assist with:

- Maintenance of security.
- Awareness of anomalous events (e.g., attacks).
- Maintenance of network health through monitoring and tuning.

## 4. SONIFICATION FOR NETWORK MONITORING

Much work has been done in applying information visualisation techniques to network data for facilitating situational awareness (e.g., see Jajodia *et al*. [7] for a recent overview). However, a particularly striking feature of the three-level model is that the first two levels — perception and comprehension — correspond directly with Pierre Schaeffer's two basic modes of musical listening, ´ecouter (hearing, the auditory equivalent of perception) and entendre (literally 'understanding', the equivalent of comprehension). Schaeffer was writing within a musical arts context but Vickers [8] demonstrated how these modes are applicable to sonification.

Sonification is a branch of auditory display, a family of representational techniques in which non-speech audio is used to convey information. Here, data relations are mapped to features of an acoustic signal which is then used by the listener to interpret the data. Sonification has been used for many different types of data analysis (see Hermann, Hunt, and Neuhoff [9] for a broad and recent treatment of the field) but one for which it seems particularly well suited is live monitoring, as would be required in situational awareness applications. The approach described in this chapter provides one way of addressing the challenges outlined above by enabling operators to monitor infrastructures concurrently with other tasks using additional senses. This increases the available bandwidth of operators without overloading individual cognitive functions, and provides a fast and elegant route to practical situational awareness using multiple senses and an increased range of cognitive ability.

Situational awareness requires intelligence to be provided in real time. A major challenge with live real-time network monitoring is that, with the exception of alarms for discrete events, the administrator needs to attend to the console screen to see what is happening. Spotting changing or emerging patterns in traffic flow would need long-term attention to be focused on the display. Therefore, sonification has been proposed as a means of providing situational awareness.

Monitoring tasks can be categorised as direct, peripheral, or serendipitous-peripheral:

> In a direct monitoring task we are directly engaged with the system being monitored and our attention is focused on the system as we take note of its state. [3, p. 455]

A system to sonify network traffic, on the other hand, would allow monitoring in a peripheral mode. Here,

> … our primary focus is elsewhere, our attention being diverted to the monitored system either on our own volition at intervals by scanning the system … or through being interrupted by an exceptional event signalled by the system itself. [3, p. 455]

Hence, the monitoring becomes a secondary task for the operator who can carry on with some other primary activity. Serendipitous-peripheral is like peripheral monitoring except that the information gained "is useful and appreciated but not strictly required or vital either to the task in hand or the overall goal" [3, p. 456]. Thus, a system to sonify network traffic may allow us to monitor the network in a peripheral mode, the monitoring becoming a secondary task for the operator who can carry on with some other primary activity. Network traffic is a prime candidate for sonification as it comprises series of temporally-related data which may be mapped naturally to sound, a temporal medium [3].

Gilfix and Crouch's PEEP system [10] is an early network sonification example but Ballora *et al*. [11]–[13] developed the idea to address situational awareness. Using an auditory model of the network packet space they produced a "nuanced soundscape in which unexpected patterns can emerge for experienced listeners". Their approach used the five-level JDL fusion model which is concerned with integrating multiple data streams such that situational awareness is enhanced (see Blasch and Plano [14]. However, Ballora *et al*. [11] noted that the high data speeds and volumes associated with computer networks can lead to unmanageable cognitive loads. They concluded:

> The combination of the text-based format commonly used in cyber security systems coupled with the high false alert rates can lead to analysts being overwhelmed and unable to ferret out real intrusions and attacks from the deluge of information. The Level 5 fusion process indicates that the HCI interface should provide access to and human control at each level of the fusion process, but the question is how to do so without overwhelming the analyst with the details.

Kimoto and Ohno [15] developed a network sonification system called 'Stetho' which uses the network traffic as source of sound, based on assumption that the sound will be useful for the network administrator. The music generated by Stetho should be comfortable as music so that changes in network status and exceptional events should be immediately noticeable. Stetho reads commands from the tcpdump packet analyser, then checks and matches them to generate corresponding MIDI events (see http://www.tcpdump.org).

InteNtion (Interactive Network Sonification) is a project targeted at mapping network activity to musical aesthetic. The SharpPCap library is used to analyse the network traffic. The resultant data are then transformed into MIDI messages and sent to synthesisers to generate dynamically mixed sounds [16].

Wolf and Fiebrink [17] designed the SonNet system to help users (artists or people have an interest in network traffic information) to easily access network traffic through a simple coding interface without requiring knowledge of Internet protocols. SonNet acts as packet-sniffing tool and network connection state analyser. It includes an object from the ChucK concurrent music programming language that can be used to generate the required audio (see http://chuck.cs.princeton.edu).

Users typically employ third party packet sniffing applications or libraries such as Wireshark, Tcpdump, or Carnivore. Users have to write code to adjust these tools in order send the network information to a sonic environment in real time. SonNet shortens the process of creating sonifications from network data. With SonNet users do not need to write code to access networking data or write code to track network state information from packet data. SonNet organises the network data into three levels of abstraction: 1) raw packet data, 2) single packet analysis, and 3) accumulated packet analysis [17]. The first level deals with the raw information contained in network packets. By analysing source and destination IP addresses and port numbers level 2 deals with determining the direction of a packet's travel (in or out of the gateway) plus the elapsed time since the previous packet. Level 3 is an aggregation of this analysis and also computes the packet rate over a user-defined time period and the running average packet rate. Thus, the three levels provide different views of the packet data.

### 5. SONIFICATION WITH SOUNDSCAPES

A major challenge for sonification designers continues to be that their work is often perceived as annoying, fatiguing, or both. Whilst annoying and fatiguing sonifications might be tolerable for short tasks, for monitoring tasks (especially those in which situational awareness is the goal) something better is needed. In these situations the environment is unlikely to be a controlled scientific laboratory in which extraneous noises can be removed. For the network administrator, especially one trying to attain situational awareness in a stressful situation such as a cyber attack, the working environment will be sonically uncontrolled. Here sonifications are needed that are not only *not* annoying or fatiguing but which complement the existing sonic environment.

Vickers, Laing, and Fairfax [4] demonstrated their Self-Organised Criticality Sonification System (SOCS) that sonifies meta properties of network traffic data using a soundscape approach (see Fig. 1). The concept of soundscape was introduced by Schafer [18] and is one form of sonic organisation which can be applied to the sonification of a network environment. Pijanowski *et al.* [19, p. 203] observed that sounds

> … are a perpetual and dynamic property of all landscapes. The sounds of vocalizing and stridulating animals and the non-biological sounds of running water and rustling wind emanate from natural landscapes. Urban landscapes, in contrast, are dominated by human-produced sounds radiating from a variety of sources, such as machines, sirens, and the friction of tires rotating on pavement.

A soundscape ecology what is formed by the sounds and spatial temporal patterns as they are created by a landscape's environment [19]. Ecology is studying the relationship between (individuals and communities) within their living environment. Therefore, soundscape ecology is studying the effects of the acoustic environment created by those living with in it due to their responses and behavioural characteristics. The impetus behind it is to recognise imbalances which may have unhealthy or malicious effects.

In principle, a well designed sonification soundscape will either fit in with the existing environment or will sit alongside it in a complementary manner. We are already used to dealing with everyday background sound and quickly deciding what sounds need attending to and what sounds can be pushed to the attentional background. A soundscape offers the sonification designer the potential to leverage this innate information processing capacity in such a way that important changes in the cyber environment become salient in the soundscape.

There have been some notable recent advances in taxonomy for sonification and its relationship to listening (see Tuuri and Eerola [21], Vickers [8], Grond and Herman [22], and Filimowicz [23]). Typically, using Schaeffer's *quatre écoutes* and Chion's causal and semantic listening [24] as a starting point, sonification listening categorisations have been put forward as tools to help in the exploration and understanding of the interactions between a listener and a sonification. For example, Vickers [8] extended Shaeffer's scheme by adding further four listening modes that pertain to sonification. Tuuri and Eerola [21] proposed an alternative three-level taxonomy with eight listening modes. Bringing together these taxonomical accounts and the ecological approach of the soundscape offers the potential to design sonifications that are effective communication channels at the same time as being environmentally compatible and less fatiguing (what Adams *et al.* [20] might call a 'sustainable soundscape'). In the study of natural soundscapes Pijanowski *et al.* state that research "is needed on how natural sounds influence the development of individuals' sense of place, place attachment, and connection to nature" [19, p. 209]. Likewise, sonification research will need to explore how soundscape sonification influences the listener's development of a sense of the information space and their own place within it.

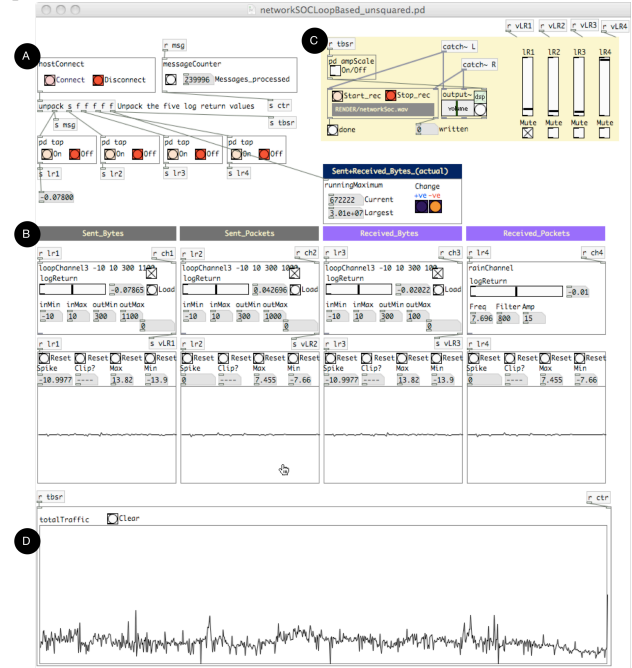

Figure 1: Screen shot of the Self-Organised Criticality Sonification System [4]. Section A receives network traffic from a capture script. Section B shows the voice definitions to which each traffic variable is mapped. Section C mixes the audio streams into a stereo feed. Section D is a combined plot of the variables being monitored.

## 6. Conclusion

Twenty years ago, Kramer called for collaboration between sonification designers and composers [25]. Despite a few notable exceptions the sonification community seems to have been unwilling or unable to enter into such collaborations (and often with justifiable reticence, for example, see Bovermann, Rohruber, and de Campo [26, p. 240]). However, recently there has been an increasing interest in exploring the aesthetic aspects of sonification (e.g., see Schedel and Worrall's editorial [27]) and the definitional boundary between sonification and music continues to be pushed by sonification designers and composers alike [28]. We now see designers on the one hand who are thinking seriously about the role of aesthetics in sonification design and composers on the other hand who are increasingly interested in using data and sonification schemata in their own aesthetic practice.

It is the goal of our present research to produce a real-time network monitoring system using a soundscape based interactive sonification to enhance situational awareness for network administrators. Such a system will enable them to monitor network activities while performing other administration tasks in order to recognise and identify any patterns of sound that indicate misuse or malicious activity to achieve real-time intelligence about a network environment. The initial SOCS prototype [4] serves as an initial proof of design concept. The next stage will bring together our work on situational awareness [1] and a more formal soundscape approach in a new tool. It is intended to experiment with a range of different naturalistic and artificial soundscapes (e.g., forest, city, sound effects) to see which works best in supporting situational awareness. Modern networks generate a lot of data and it is hoped that a soundscape approach will offer an environmentally complementary solution that is acceptable to users and which minimises annoyance and fatigue.